\documentclass[fleqn,usenatbib]{mnras}

\usepackage{newtxtext,newtxmath}
\usepackage{lipsum}
\usepackage[T1]{fontenc}
\usepackage{ae,aecompl}

\usepackage[utf8]{inputenc}
\usepackage{graphicx}	
\usepackage{amsmath}	

\newcommand{\bb}{{\mathbf b}}

\newcommand{\bbf}{{\mathbf f}}
\newcommand{\bx}{{\mathbf x}}

\newcommand{\bX}{{\mathbf X}}
\newcommand{\bY}{{\mathbf Y}}
\newcommand{\bA}{{\sf A}}
\newcommand{\C}{{\sf C}}

\newcommand{\LL}{{\sf L}}
\newcommand{\cL}{{\cal L}}
\newcommand{\V}{{\sf V}}
\newcommand{\W}{{\sf W}}
\newcommand{\X}{{\sf X}}
\newcommand{\Z}{{\sf Z}}

\newcommand{\bu}{{\bf u}}
\newcommand{\bv}{{\bf v}}
\newcommand{\bbmu}{\mbox{\boldmath $\mu$}}

\title{Extreme data compression while searching for new physics}

\author[A. F. Heavens et al.]{
Alan F. Heavens$^{1}$\thanks{E-mail: a.heavens@imperial.ac.uk},
Elena Sellentin$^{2}$,
and Andrew H. Jaffe$^{1}$
\\
$^{1}$Imperial Centre for Inference and Cosmology (ICIC), Imperial College, Blackett Laboratory, Prince Consort Road, London SW7 2AZ, U.K.\\
$^{2}$Leiden Observatory, Leiden University, Huygens Laboratory, Niels Bohrweg 2, NL-2333 CA Leiden, The Netherlands.
}

\date{Accepted 5 May 1944. Received 5 August 1888; in original form 3 May 1849}

\pubyear{2020}

\begin{document}
\label{firstpage}
\pagerange{\pageref{firstpage}--\pageref{lastpage}}

\maketitle

\begin{abstract}
Bringing a high-dimensional dataset into science-ready shape is a formidable challenge that often necessitates data compression. Compression has accordingly become a key consideration for contemporary cosmology, affecting public data releases, and reanalyses searching for new physics. However, data compression optimized for a particular model can suppress signs of new physics, or even remove them altogether. We therefore provide a solution for exploring new physics \emph{during} data compression. In particular, we store additional agnostic compressed data points, selected to enable precise constraints of non-standard physics at a later date.
Our procedure is based on the maximal compression of the  MOPED algorithm, which optimally filters the data with respect to a baseline model. We select additional filters, based on a generalised principal component analysis, which are carefully constructed to scout for new physics at high precision and speed.
We refer to the augmented set of filters as MOPED-PC.  They enable an analytic computation of Bayesian evidences that may indicate the presence of new physics, and fast analytic estimates of best-fitting parameters when adopting a specific non-standard theory, without further expensive MCMC analysis. As there may be large numbers of non-standard theories, the speed of the method becomes essential. Should no new physics be found, then our approach preserves the precision of the standard parameters. As a result, we achieve very rapid and maximally precise constraints of standard and non-standard physics, with a technique that scales well to large dimensional datasets.
\end{abstract}

\section{Introduction}
It may seem obvious that more data will lead to better constraints on the free parameters of a physical theory from a data set. In reality, however, there may be considerable scope to reduce the size of the dataset without compromising precision on the parameters. Data compression can be a valuable tool, especially in the cosmological context of large datasets, when relatively few parameters are to be inferred, and may be necessary, for likelihood-free inference \citep{Alsing, Leclercq} or when data covariance matrices have to be estimated numerically \citep{Heavens2017}.  

The possibility of radical data compression is easily evidenced by Gaussian-distributed data points whose expectation value $\mathbf{\mu}$ only is to be inferred: if there are $N$ data points, $x_1,...,x_n$, each drawn from the same Gaussian distribution, then these can be compressed to a single datum $\bar{x} = \sum_i x_i/N$, which contains the same information on $\mathbf{\mu}$ as the original $N$ points. Data compression which conserves the data's information content may be said to be optimal or lossless, and lossy if information is lost during the compression\footnote{Lossless here has a restricted meaning of no loss of precision in the parameters, and is a much weaker demand than requiring that the original dataset can be reconstructed from the compressed data.}.  The MOPED algorithm \citep{HJL} draws on a physical theory to induce such a lossless data compression. It uses the theory to radically compress the original data from their original size down to the number of parameters of the physical theory. 

However, what if the physical theory used by MOPED to compress the data is not the correct one? It then runs the risk of `compressing away' the information needed to study extended theories.  In this paper we therefore tackle the challenge of inferring physics from compressed data sets whose original uncompressed data are much larger than the number of free parameters of the theory.  We target jointly the most precise constraints on the physical parameters as possible, alongside studying whether the entire theory is supported by the data or not. It might well be possible that for no value of the free parameters does the theory fit the data satisfactorily, and this may motivate the search for alternative theories.  It is important to realize that the stage of data compression might be the last instance where a challenge to the theory is possible: the compression may suppress, or even remove, signals of new physics, as it only strives to preserve the information content of the standard theory.
 
Additionally, a computational bottleneck arises: if a theory is abandoned, a replacement is needed to explain the data. If there exist no compelling theoretical reasons which single out an unambiguous replacement theory, then iterative testing over a huge theory space begins. This has, e.g., been evidenced by the track record of cosmology, where a discrepancy between the posteriors of the Planck satellite \citep{Planck2018} and CFHTLenS \citep{LensingLow} triggered a decade-lasting series of proposed theory amendments, which continues to last into the era of the Kilo Degree Survey \citep[KiDS;][]{KV450} and the Dark Energy Survey \citep[DES;][]{DES}. 

With the launch of ESA's Euclid satellite \citep{Euclid}, and the
`Legacy Survey of Space and Time' \citep[LSST;][]{LSST} starting operations in the near future, a continuing trend of proposing and testing non-standard theories is foreseeable. In light of the rapidly increasing data size, non-strategic testing of theories is however rapidly becoming numerically prohibitive. To address this, we provide a principled, largely analytical, strategy of testing whether a data set shows any evidence of being incompatible with a standard theory, while also compressing it such that it maintains its information content, and facilitating inference of standard and non-standard physics. Both the analytical solutions and the compression are anticipated to facilitate reanalyses of Euclid- and LSST-sized datasets upon their public releases.

We base our study on the MOPED (`Massive Optimised Parameter Estimation and Data compression') extreme data compression algorithm, published in \citet{HJL} and \citet{RJH}. MOPED is a linear compression algorithm, which sifts the initial data through a set of filters, whereupon the filters output compressed data points. MOPED constructs these filters by assuming the data are well fitted by a parametric theory with $M$ free parameters. It compresses from initially $N$ data points down to $M$ data points, i.e. one datum per parameter. It is optimal in the sense that the Fisher matrix of the compressed data is identical to the Fisher matrix of the original dataset in the case that 1) the data are Gaussian-distributed, 2) the information comes from the expectation value of the data, not its covariance, 3) the fiducial parameter set (used to define the filter functions) is the correct set of values.   Condition (3) is clearly not possible to guarantee a priori, but one can perform a sub-optimal analysis based on an initial guess, determine maximum-likelihood parameters, and compress again with these as the fiducial model, iterating as needed.  Practice has found extremely rapid convergence, and often no iteration at all is required \citep{HJL}.

This approach is enormously powerful if the theory used for compression is the correct model for the data.  The purpose of this paper is to include an investigation of whether the data to be compressed actually indicate any evidence of disfavouring the compressing theory. If the data show no sign of disfavouring the theory, then MOPED is complete.

What happens if the theory requires extension? MOPED by design produces too few data points to test alternative theories. After compression, falsifying the theory is extremely difficult, as the information incompatible with the theory will largely have been dropped during compression. Likewise, due to having compressed down to $M$ data points, a theory with more than $M$ parameters is difficult to constrain. If the dataset has been compressed in order to ease its public release, then the compression to $M$ data points directly hinders public reanalyses to study extended models.

In this paper, we therefore define a set of additional compression vectors - basis functions that can investigate extensions of the baseline physics model. Any number of these components can be added, up to the size of the original dataset, and the amplitudes of these modes represent additional parameters to be determined.  Although allowing extra flexibility, the price seems to be a potentially large increase in the dimension of the parameter space and a consequently very challenging sampling problem.  However, by carefully constructing these modes, we can define a linear model for the extra parameters, based on a generalised principal component analysis (gPCA).  The posterior distribution of the additional parameters can then be determined analytically, with almost no additional computational cost.  Bayes factors can be computed analytically to determine if an extended model is favoured, and the highest amplitude modes can indicate in which data directions the model needs extending.  Furthermore,  extended physical models can be expanded in this basis set, allowing their parameters to be inferred at minimal extra cost, and in particular without performing a higher-dimensional Monte Carlo Markov Chain (MCMC) sampling. Our technique allows not only extreme compression of the data, but simultaneously permits model comparison, and additionally creates analytical shortcuts which replace otherwise numerical brute-force approaches in inference.

Throughout the paper, we will refer to the theory on which the MOPED compression is based as the {\em baseline physics model}. We refer to our extension of the baseline theory, in the gPCA modes, for falsification purposes as the {\em agnostically extended model}. The extension is agnostic as it introducing the \emph{maximum} number of possible additional parameters, and may be used to test the correctness of the baseline physics model. This provides the most complete study of deviations from the baseline theory. Finally, we can linearly expand {\em specific extended models} in the agnostic extension basis set. These specific models may be physically or astronomically motivated and introduce fewer extra parameters to capture potential deviations from the standard theory. These models can be tested analytically against the baseline model using Bayesian model comparison techniques.  As mentioned, all of this can be achieved via simple matrix operations, without running more, potentially very challenging, MCMC chains. Finally, our agnostic parameters are introduced such that they do not weaken the constraints on the primary parameters of the baseline physical model: this is of central importance, as the introduction of additional nuisance parameters usually inflates the uncertainty of the parameters of primary interest.

The layout of the paper is as follows. In section~\ref{sect:MOPED} we review the MOPED radical data compression algorithm, which is designed around a particular (baseline) model. In section~\ref{sect:gPCA} we introduce a basis of additional orthogonal vectors as `agnostic' extensions to the baseline theory, and show that the mode amplitudes can be determined analytically without further MCMC studies.  In section~\ref{sect:Models_specific} we show how specific physics-driven extensions to the baseline theory may be tested analytically, by projecting their signal derivatives onto the extended modes. In section~\ref{sect:GalaxyImages} we consider a detailed illustrative example of determining structural parameters from a galaxy image
and explore compressing the extended modes to a subset of those that are most discrepant with the theory. We show that retaining even a few percent of modes may be sufficient to challenge the baseline model and constrain its competitors. In total, this implies public reanalyses of Euclid- or LSST-sized data do not need to succeed in analyzing the full unwieldy data sets, but may study new physics from much smaller, carefully prepared, compressed data sets. Finally, in section~\ref{sect:conclusions} we present our conclusions.

\section{MOPED with unchallenged theories}
\label{sect:MOPED}
We first review MOPED data compression.  Introduced by \cite{HJL}, it is a linear compression method that is optimised for Gaussian-distributed data of fixed covariance matrix.  The $N$-dimensional data vector $\bx$ is drawn as
\begin{equation}
    \bx \sim {\cal N}(\bbmu,\C),
\end{equation} 
where $\mathcal{N}$ is the multivariate Gaussian distribution of mean $\bbmu$ and covariance matrix $\C$. Denoting expectation values of the data with angled brackets, it is now conjectured that there exists a parametric prediction for the expectation value, e.g.~a model from theoretical physics which might explain the data. Assuming this model has $M$ free parameters, we then have
\begin{equation}
    \langle\bx\rangle \equiv \bbmu(\boldsymbol{\Theta}),
\end{equation}
where $\boldsymbol{\Theta} = (\Theta_1,..., \Theta_M)$ are the $M$ free parameters of the theory. In a usual inference problem, the values of these parameters are to be inferred.

We assume that the expectation value $\bbmu(\boldsymbol{\Theta})$ is differentiable at least twice with respect to its parameters, and write $\partial \bbmu/\partial \Theta_\alpha = \bbmu,_\alpha$, where the minor Greek index $\alpha$ here runs over the $M$ free parameters of the theory, i.e.~$\alpha \in[1,M]$. Then, the MOPED algorithm defines one MOPED weighting vector $\bb_\alpha$ per free parameter of the theory. These MOPED vectors act as filters when applied to the data. The dot products of the MOPED filter vectors with the original data set $\bx$ thus produce $M$ compressed data points
\begin{equation}
  y_\alpha = \bb_\alpha^T \bx, \ \alpha \in [1,M].
  \label{y}
\end{equation}
The MOPED filters $\bb_\alpha$ have been derived in \cite{HJL}, and follow from an optimization procedure, such that the set of compressed data points $\{y_\alpha\}$ contains the same Fisher information on the model parameters $\boldsymbol{\Theta}$ as the full data set $\bx$. Under this optimization criterion, the weighting vectors were shown by \citep{HJL} to be
\begin{equation}
\bb_1 = \frac{\C^{-1} \bbmu_{,1}}{\sqrt{\bbmu_{,1}^T
\C^{-1}\bbmu_{,1}}}, \ \ \ \
\bb_{\alpha>1} = \frac{\C^{-1}\bbmu_{,\alpha} - \sum_{\beta=1}^{\alpha-1}(\bbmu_{,\alpha}^T 
\bb_\beta)\bb_\beta}{
\sqrt{\bbmu_{,\alpha}^T \C^{-1} \bbmu_{,\alpha} - \sum_{\beta=1}^{\alpha-1}
(\bbmu_{,\alpha}^T \bb_\beta)^2}}.
\label{MOPED1}
\end{equation}
For $M>1$ the modes are Gram-Schmidt orthogonalized with respect to the covariance matrix as metric tensor. This is an optional step which does not interfere with optimality, and one can also use simpler non-orthonormalised weights such as $\C^{-1}\bbmu_\alpha$, as used by \cite{Gualdi} and \cite{ZablockiDodelson}, and the compression is equivalent to MOPED.  To set the weights, a fiducial parameter set needs to be defined, at which the derivatives are computed.

The Gram-Schmidt procedure ensures orthonormality of the modes, which is defined by the relation
\begin{equation}
\bb_\alpha^T \C \bb_\beta = \delta_{\alpha\beta}.
\end{equation}
With Gram-Schmidt orthogonalisation, the ultimate outcome is that the compressed data points of Eq.~(\ref{y}) are uncorrelated, and each has unit variance. i.e. the covariance matrix of the data points $\{y_\alpha\}$ is thus by construction the $M \times M$ identity matrix.  Note that this only holds approximately if the fiducial model is not correct, and this is explored further in \cite{Heavens2017}, which also treats parameter-dependent covariance matrices: data compression facilitates covariance estimation, and in particular inversion, if the covariance matrices are estimated from simulations \citep{SH,SH17}.  The MOPED compression has been extended to nonlinear compression by \cite{AlsingWandelt18}, who provided further insight into why it works, connecting it to the $M$ components of the gradient of the likelihood function.  
In the same spirit, neural networks can be used for non-Gaussian data distributions to find the optimal weights \citep{Charnock}.  Finally, \cite{ZablockiDodelson} formed linear combinations of the MOPED coefficients to remove some parameter dependence, at least locally. 

In practice, the compression can be massive, since it reduces the original dataset of dimension $N$ to the length of the parameter vector, which is $M$. The compressed data set then contains one data point per parameter to be measured.
MOPED is exactly preserving the Fisher information if the derivatives of the weighting vectors are evaluated at the correct value of the model parameters. For most applications the Fisher matrix will however vary little throughout parameter space, such that MOPED compression is almost optimal also in a realistic setting where the correct parameter values are not known a priori. In total, MOPED thus can lose no information in the Fisher sense, despite its very radical compression. However, 
a limitation is that the compression is tied to a particular theory, and does not lend itself naturally to the comparison of different theories.

\section{Data compression with model comparison}
\label{sect:gPCA}

\subsection{Adding additional filters}
With $N$ data points and $M$ theoretical parameters used by the baseline physical model, there are $M$ MOPED vectors, and $N-M$ additional degrees of freedom that could be explored. We thus use these additional degrees of freedom to introduce auxiliary parameters, for the purpose of testing whether the baseline physical model is sufficient to explain the data, or whether a theory with extended parameter space is required.

Contrasting two theories with each other, and computing which of the two is preferred by the data, is answered by Bayesian model comparison. However, introducing new parameters in an arbitrary way will compromise the precision with which $\boldsymbol{\Theta}$, the parameters of the baseline physical model, are measured. If more parameters are to be jointly inferred, then the uncertainty on each of them will generally increase. If we do not wish to compromise the precision of the baseline physical parameters, then the only possibility is to introduce new parameters under the additional constraint that these be statistically \emph{independent}.

Enforcing independence between parameters can, at the level of a Gaussian approximation, be ensured if the extended model is linear in the new parameters, and if the Fisher matrix between old and new parameters is a block-diagonal matrix, as the correlation between the old and new parameters is then zero. To meet these goals, we thus introduce the additional parameters in the following manner.

The baseline physical model will restrict the expected data vector to an $M$-dimensional hypersurface in the $N$-dimensional data space.  The actual data vector will not lie in this hypersurface, either due to noise, and/or because the physical model is not correct.  Our task is to determine which. 

Let us consider the part of the data vector $\bx$ that is not in the baseline model hypersurface. This part results from removing the projections of the original data vector onto the MOPED vectors, locally at $\bbmu_F \equiv \bbmu(\Theta_F)$, where $\Theta_F$ is the fiducial set of baseline parameters.  For convenience, we subtract this from $\bx$:
\begin{equation}
\label{eq:project}
\bX \equiv (\bx-\bbmu_F) - \sum_{\alpha=1}^M \left[\bb^T_\alpha  (\bx-\bbmu_F) \right] \C \bb_\alpha.
\end{equation}
The resultant vector $\bX$ effectively points orthogonally out of the hypersurface at the point $\bbmu_F$.  With this definition, $\bX$ is orthogonal to all the MOPED vectors, defined at $\boldsymbol{\Theta}_F$, in the normal sense, i.e. $\bb_\alpha^T \bX = 0$ for all $\alpha \in [1,M]$.  It then follows that the covariance matrix of $\bX$ is given by
\begin{equation}
\C^X_{ij} = \C_{ij} - \sum_{\alpha=1}^M (\C\bb_{\alpha})_i ( \C \bb_\alpha)_j.
\label{CX}
\end{equation}
The Roman indices $i,j$ label the different vector or matrix elements, whereas the Greek index $\alpha$ runs over the parameters.

\subsection{Agnostic extension of the baseline theory}

The procedure that we follow is to define a set of basis vectors $\bu_\gamma; \gamma \in [M+1,N]$, that span the $N-M$-dimensional space orthogonal to the baseline model hypersurface at the fiducial model $\bbmu_F$.  Deviations from the baseline model will be reflected in non-zero amplitudes of these modes, but $\bX$ will
also be non-zero because of noise.  Our aim is to define modes whose amplitudes can be computed analytically -- this frees us from otherwise having to perform an $N-$dimensional MCMC analysis, which could be prohibitively expensive. We can avoid the MCMC by constructing a linear data model that is effectively a Taylor expansion of the expected signal, adding extra degrees of freedom corresponding to directions in which the expected data vector can move, so the expected signal is modified to 
\begin{equation}
\bbmu(\Theta,\Gamma) = \bbmu(\boldsymbol{\Theta}) +\sum_{\gamma=M+1}^{N} \Gamma_\gamma \C \bu_{\gamma},
\label{mugen}
\end{equation}
where we have expressed as $\C\bu_\gamma$ the partial derivative of the signal with respect to the new parameter $\Gamma_\gamma$.  The first term on the right hand side is $\bbmu(\boldsymbol{\Theta},\boldsymbol{\Gamma}=\boldsymbol{0})$, which is just the expected signal $\bbmu(\boldsymbol{\Theta})$ for the baseline physics model. This first-order Taylor expansion in the directions off the hypersurface yields linear parameters in the model that can be treated analytically.  

To find the new modes $\bu_\gamma$, it is convenient if they have the same orthonormality properties as the MOPED vectors, which span the hyperplane locally.  i.e. they are normal to each other,
\begin{equation}
\bu_\alpha^T \C \bu_\beta = \delta_{\alpha\beta},\qquad \alpha,\beta \in [M+1,N]
\end{equation} 
and also to the MOPED vectors,
\begin{equation}
 \bb_\alpha^T \C \bu_\gamma = 0; \qquad \alpha \in [1,M], \gamma \in [M+1,N].
\end{equation}

We can satisfy these constraints by solving the generalised eigenvalue problem
\begin{equation}
\C^X \bu_\gamma = \Lambda_\gamma \C \bu_\gamma.
\label{ortho}
\end{equation}

The resulting $N-M$ generalized eigenvectors $\bu_\gamma$ with non-zero eigenvalues will form our additional generalised PCA (gPCA) weighting vectors.  As many of these as desired are then added to the set of MOPED vectors.

To solve the generalised eigenvalue problem of Eq.~(\ref{ortho}), we Cholesky decompose the positive definite covariance matrix $\C = \LL \LL^T$. Rearranging, and introducing the shorthand  $\LL^{-T} = (\LL^T)^{-1}$, this leads to
\begin{equation}
(\LL^{-1}\C^X \LL^{-T}) \LL^T \bu_\gamma = \Lambda_\gamma \LL^T \bu_\gamma,
\end{equation}
which is now an ordinary eigenvalue problem
\begin{equation}
    \bA \bv_\gamma = \Lambda_\gamma \bv_\gamma,
\end{equation} 
where the matrix $\bA \equiv \LL^{-1}\C^X \LL^{-T}$, and
\begin{equation}
   \bv_\gamma = \LL^T \bu_\gamma, 
   \label{identifyu}
\end{equation}
are the eigenvectors of $\bA$ with eigenvalues $\Lambda_\gamma$.
We can find our generalized eigenvectors $\bu_\gamma$  by solving  Eq.~(\ref{identifyu}).

The eigenvectors $\bv_\gamma$ can be made orthogonal in the usual sense, and indeed have to be, since it is straightforward from the expression for $\C^X$ (equation \ref{CX}) to show that the eigenvalues of the problem are either 0 (with multiplicity $M$), or 1 (with multiplicity $N-M$).  The off-hypersurface eigenvectors (with eigenvalue 1) are arbitrary provided they are orthogonal, but their precise directions are not important -- any mutually orthogonal set will do.  The $M$ eigenvectors $\bu_\gamma = \LL^{-T}\bv_\gamma$ with eigenvalues equal to 0 are discarded, as they lie in the local hyperplane already spanned by the MOPED vectors, and the data dependence there is already captured by the MOPED vectors.

From the (usual) orthogonality of $\bv_\gamma$, the orthogonality condition for the $\bu_\gamma$ is 
\begin{equation}
 \delta_{\alpha\beta}=\bv_\alpha^T \bv_\beta = (\LL^T \bu)_\alpha^T  (\LL^T \bu)_\beta =
\bu_\alpha ^T \C \bu_\beta,
\end{equation}
as desired.
We then add the gPCA $\bu_\gamma$ eigenvectors for which $\Lambda_\gamma =1$ (i.e. $\gamma \in [M+1,N]$  to the set of MOPED filters $\bb_\alpha$. 

The filters with which we now sift through the data are then
\begin{equation}
  \mathbf{f}_\alpha=
  \begin{cases}
    \bb_\alpha, & \text{for $1\leq \alpha\leq M$},\\
    \bu_\alpha, & \text{for $M+1\leq \alpha\leq N$ }.
  \end{cases}
\end{equation}
and the data points are
\begin{equation}
    Y_\alpha = \mathbf{f}_\alpha^T (\bx-\bbmu_F), \ \ \ \alpha \in [1,N],
\end{equation}
where we have for convenience modified the MOPED algorithm to subtract the expected signal at the fiducial point.

The first $M$ data points, corresponding to the MOPED modes, give the maximum useful compression if the theory is correct.
The remaining $N-M$ data points serve the purpose of being able to falsify the baseline physical theory. If the baseline theory is correct, then these data points contain no information, only noise. If the dataset is to be published for reanalysis, then the $M$ MOPED datapoints would be provided, and an `interesting' subset (to be defined in Sect.~\ref{FMDC}) of the additional points should be added, to enable model comparison and studies of non-standard theories with the public dataset. 

This establishes our complete set of weight vectors.  Of course, if we choose to keep all $N-M$ gPCAs, then there is no compression at all, but we can combine the advantages of massive data compression with an investigation of whether the physical model requires extension by including a subset of the extra modes.  As we will see in the next sections, this construction has a number of very useful properties.

\subsection{Extended model parameters}

In this section, we derive the posterior distribution of the extended model parameters, $\Gamma_\gamma$, introduced in Eq.~(\ref{mugen}). We will show that, subject to certain benign conditions, their marginal posteriors are independent Gaussian distributions of unit variance. The mode of their posterior can be computed analytically, and they fully decouple from the baseline model parameters; introducing $N-M$ additional parameters $\Gamma_\gamma$ for the sake of model comparison thus does not deteriorate our knowledge of the primary physical parameters.

The analytic solution is an important practical point:  We may infer, if desired, all $N-M$ extra parameters without undertaking an expensive $N-$dimensional MCMC sampling.

\subsubsection{Gaussianity}

If the original data $\bx$ are Gaussian-distributed, then so are the compressed data, since they are linear combinations of $\bx$.  If we assume a Gaussian prior on $\Gamma$ (a uniform distribution also works), then the posterior probability as a function of $\Gamma_\gamma$ will be Gaussian, since
\begin{equation}
    \mathcal{P}(\Gamma, \Theta | \bY) \propto {\mathcal P}(\bY | \Theta, \Gamma) \pi(\Theta, \Gamma)
\end{equation}
where $\pi$ is the prior, Gaussian in $\Gamma$, and the likelihood $\mathcal{P}(\bY | \Theta, \Gamma)$ is Gaussian in both $\bY$ and $\Gamma$:
\begin{equation}
    \mathcal{P}(\bY | \Theta, \Gamma) = (2\pi)^{-N/2}\exp\left(-\frac{1}{2}
    \sum_{\beta=1}^N \left\{Y_\beta - \bbf^T_\beta\left[\bbmu(\Theta,\Gamma)
    -\bbmu_F\right]\right\}^2\right),
\end{equation}
where $\bbmu(\Theta,\Gamma)$ is linear in $\Gamma_\gamma$ and given by equation (\ref{mugen}).

The maximum of the likelihood (ML) is at $(\Theta_{\rm ML},\Gamma_{\rm ML})$, where $\Gamma_{\rm ML}$ can be computed analytically for given $\Theta_{\rm ML}$, since $\bbf$ and $\bu$ are orthogonal:
\begin{equation}
    \Gamma_{\rm ML,\gamma} = Y_\gamma - \bbf_\gamma^T[\bbmu(\Theta_{ML})-\bbmu_F]; \qquad \gamma \in [M+1,N].
    \label{GammaML}
\end{equation}
This is an important result.  If the prior on $\Gamma_\gamma$ is uniform, then the peak of the posterior is given analytically by equation (\ref{GammaML}), and this is easily generalised to the peak of the posterior when a Gaussian prior is assumed.  So we find that once a MOPED analysis has been done to find $\Theta_{\rm ML}$, only trivial computations are required to infer $\Gamma_\gamma$.

\subsection{Hessian matrix and data compression}
\label{FMDC}

The likelihood of the $\Gamma_\gamma$ is Gaussian, with the mean given by equation (\ref{GammaML}).  The covariance is obtained from the Hessian matrix, which is composed of four blocks, whose expressions simplify because of the orthonormality of the $\bbf_\beta$.  The $\Theta$-$\Theta$ block is the same as in the MOPED analysis, given at the peak by
\begin{equation}
    -\frac{\partial^2 \ln {\cal L}}{\partial \Theta_\alpha \partial\Theta_\beta} = \sum_{\rho=1}^N (\bbf_\rho^T\bbmu_{,\alpha})(\bbf_\rho^T\bbmu_{,\beta}).
\end{equation}
The gPCA-gPCA block of additional modes is the identity matrix in all circumstances:
\begin{equation}
    -\frac{\partial^2 \ln {\cal L}}{\partial \Gamma_\gamma \partial \Gamma_\rho} = \delta_{\gamma\rho}.
\end{equation}
The MOPED-gPCA off-diagonal block is
\begin{equation}
    -\frac{\partial^2 \ln {\cal L}}{\partial \Theta_\alpha \partial \Gamma_\gamma} = \bbf^T_{\gamma}\bbmu_{,\alpha}.
\end{equation}
Here we see that the off-diagonal blocks are zero, provided that the ML point is the fiducial model, or the baseline physics model is linear in the parameters.  Otherwise the tangent hyperplane spanned by the $\C\bb_\alpha$, which is defined at the fiducial parameters, is not tangent at the ML point.  

As a result, if the ML point is not known from a previous uncompressed analysis, then the MOPED analysis may need to be repeated, first to find the ML point, and then compressed during the second time with the found ML point as the fiducial model parameters.  In this way, the baseline physical model parameters and the extended parameters $\Gamma_\gamma$ are made independent, and the joint distribution of $(\Theta,\Gamma)$ is
\begin{equation}
    \mathcal{P}(\boldsymbol{\Theta},\boldsymbol{\Gamma}|\bx) = \mathcal{P}(\boldsymbol{\Theta}|\boldsymbol{y})\prod_{\gamma=M+1}^N\,\mathcal{N}(\Gamma_\gamma|\Gamma_{{\rm ML},\gamma},1),
\end{equation}
where the first term is the MOPED posterior.

This construction has a number of useful consequences.  Firstly, the introduction of the extra parameters has no effect on the inference of the baseline model parameters, whose credible regions are not degraded by their inclusion. Secondly, any number of the $N-M$ possible additional modes may be included.  Leaving any out is equivalent to marginalising over them.  Thirdly, in the baseline model, the expected values of $\Gamma_\gamma$ are zero and their variance unity (for uniform priors), so the means of the  posterior (in the more general case) can be used to assess by analytic model comparison techniques whether extension to the baseline model should be considered.  Finally, if we wish to retain a compressed dataset for comparing future models with the baseline model, then a sensible ordering is to keep those that are most discrepant with baseline model (those with the largest $|Y_\beta|$).

In conclusion for this section, we find that with the generalised PCA approach, the likelihood is separable, with the gPCA parameters being independent of each other and of the baseline physical model parameters used in the MOPED part.  If the derivatives are evaluated at the ML point, the expected likelihood of the gPCA parameters is the product of zero mean, unit variance Gaussians.

\subsection{Bayes factor: the need for an extended model?}
In this section we describe how to quantify whether the compressing baseline theory is insufficient and whether a new model should accordingly be sought. If the baseline is found to be insufficient, then Sect.~\ref{sect:Models_specific}. describes how a subset of the auxiliary parameters $\Gamma_\gamma$ can be used to quickly, and analytically, predict which other theory is likely to alleviate the defects of the baseline theory.

We quantify the insufficiency of the baseline theory by the Bayesian evidence: the evidence is the probability of the observed data given a model, independent of the values its free parameters might take. We denote the extended MOPED+gPCA model by $M_{\rm E}$ and the original baseline physical model by $M_{\rm B}$. We then compute the Bayesian evidence $Z_{\rm E}$ for the extended MOPED+gPCA model $M_{\rm E}$ (with as many extra gPCA parameters as desired) and the  baseline physical model, $M_{\rm B}$. These are combined to the Bayes factor, $B_{\rm EB} = Z_{\rm E}/Z_{\rm B}$, which gives the probabilistic ratio of whether the agnostically extended theory is preferred ($B_{\rm EB} > 1$), or whether the physical baseline theory is favoured ($B_{\rm EB} < 1$), assuming equal prior model probabilities.  For concreteness, we here assume all gPCAs are included, so there is no data compression, but fewer may be considered if desired. This would simply correspond to truncating the following sums early.

The evidence for the extended MOPED+gPCA model is
\begin{equation}
    Z_{\rm E} = \int \mathcal{P}(\bY | \boldsymbol{\Theta},\boldsymbol{\Gamma},M_{\rm E}) \, \pi(\boldsymbol{\Theta},\boldsymbol{\Gamma} | M_{\rm E}) \,\mathrm{d}^M\boldsymbol{\Theta}\, \mathrm{d}^{N-M}\boldsymbol{\Gamma},
\end{equation}
where $\pi$ is the prior.  For the baseline model,
\begin{equation}
    Z_{\rm B} = \int \mathcal{P}(\bY | \boldsymbol{\Theta},M_{\rm B}) \, \pi(\boldsymbol{\Theta} | M_{\rm B}) \,\mathrm{d}^M\boldsymbol{\Theta}.
\end{equation}
Since the models are nested (the baseline model corresponds to the extended model with $\Gamma_\gamma=0$), the Bayes factor is the Savage-Dickey density ratio  (see \cite{Trotta2007} for details):
\begin{equation}
    B_{\rm BE} = \frac{\mathcal{P}(\boldsymbol{\Gamma}=\mathbf{0} | \bY, M_{\rm E})}{\pi(\boldsymbol{\Gamma}=\mathbf{0} | M_{\rm E})},
\end{equation}
where we have assumed that the prior is separable, and the prior for the common parameters is the same for both models, $\pi(\boldsymbol{\Theta}|M_{\rm E})=\pi(\boldsymbol{\Theta}|M_{\rm B})$:
\begin{equation}
    \pi(\boldsymbol{\Theta},\boldsymbol{\Gamma} | M_{\rm E})  = \pi(\boldsymbol{\Theta}|M_{\rm B})\,\pi(\boldsymbol{\Gamma} | M_{\rm E}).
\end{equation}

Now we assume an uncorrelated Gaussian prior centred on zero for the extended model parameters
\begin{equation}
\pi(\boldsymbol{\Gamma} | M_{\rm E}) = \frac{1}{\prod_\gamma \sqrt{2\pi}\Sigma_\gamma}\,\exp\left(-\sum_{\gamma=M+1}^N  \frac{\Gamma_\gamma^2}{2\Sigma_\gamma^2}\right).
\end{equation}
Performing the Gaussian integrals to normalise properly the posterior in the numerator, we find an analytic expression for the Bayes factor

\begin{equation}
 \ln B_{\rm BE} = 
 \frac{1}{2}\sum_{\gamma=M+1}^N \left[ \ln(1+\Sigma^2_\gamma) - \frac{\Gamma^2_{{\rm ML},\gamma}}{1+ \Sigma_\gamma^{-2}}\right].
\end{equation}
There is no obvious way to set $\Sigma_\gamma$, which we will assume to be the same for all $\gamma$, but the problem does have a natural scale, since by construction the expected width of the likelihood for these components in the baseline model is unity.  It makes sensible to choose a prior that is greater than unity, and  for this paper, we have typically chosen $\Sigma_\gamma=5$, but quote some results with a prior width of 10.

However, it enters only logarithmically if $\Sigma_\gamma \gg 1$.  The sum is over those extra gPCA parameters that are included.  The greater the values of $\Gamma^2_{{\rm ML},\gamma}$, the more the gPCA model is favoured, i.e. the less adequate is the original baseline theory.

For completeness, we note that the $\chi^2$ 
of the best-fit base model may be computed in this basis as
\begin{equation}
    \chi^2 = \sum_\gamma \Gamma^2_{{\rm ML},\gamma}.
\end{equation}

In summary, our setup allows to compute the Bayes factors for agnostically extended models analytically, which avoids the otherwise highly non-trivial numerical integration to yield an evidence. This therefore removes an important computational bottleneck. Since inferring the amplitudes of additional modes is almost certainly much faster than performing the baseline MCMC analysis, there is little reason not to infer all of the gPCA mode amplitudes.

Finally, if we assume the evidence calculation indicates that the baseline model is indeed insufficient, then we now require a fast route to establishing which other physically motivated theory will fix the flaws of the baseline theory. This will be solved in the following Sect.~\ref{sect:Models_specific}.

\section{Parameter inference with specific extended models}
\label{sect:Models_specific}
One advantage of constructing our basis of agnostic extended models is that we can easily investigate {\em specific extended models} by projecting them on to the gPCA eigenmodes. This again enables us to constrain the parameters of extended theories analytically, alongside testing whether these extended theories solve shortcomings of the baseline theory. Both these issues would otherwise need to be investigated numerically.

The caveat here is that we will use a first-order Taylor expansion, so a fiducial set of extended parameters needs to be assumed. If the derivatives $\bbmu_{,\phi}$ with respect to the extended model parameters $\phi$, are strongly parameter-dependent, it may be necessary to iterate to a solution in the extended model space, and perform the linear expansion there.  Fortunately this can all be done analytically without higher-dimensional MCMC chains in the extended parameter space.

We thus assume there exists one, or multiple, extended theories which use the physically motivated parameters $\Phi_\phi$ with $\phi=1\ldots s$. We assume here that the baseline theory is `nested' into the extended theory, i.e. for $\Phi = 0$ the extended model\footnote{We assume w.l.o.g. that the baseline model has $\Phi_\phi\equiv 0$.} falls back onto the baseline theory. Although this is not strictly necessary, it simplifies the mathematical formalism below.

Constraining the extended theory implies the posterior distribution of the new parameters $\Phi$ must be calculated. Additionally, the introduction of new parameters may also shift the ML values of the former parameters $\Theta$. We thus also have to compute the shifts  $\Delta\Theta_\alpha; \alpha=1\ldots m$ of the baseline model parameters arising from the shift in $\Phi_\phi$ away from zero.

With $\Theta$ representing the baseline parameters, $\Gamma$ the agnostic extension parameters, $\Phi$ the specific extended model parameters, and $\bar Y_\beta$ being the mean predicted by the model, the log-likelihood reads
\begin{equation}
-2\ln\cL = \sum_{\beta=1}^n [Y_\beta - \bar Y_\beta(\Theta, \Gamma\  {\rm or\ } \Phi)]^2.
\end{equation}
Treating $\cL$ as a function of $\Theta, \Phi$, the ML is at 
\begin{eqnarray}
\frac{\partial \ln\cL}{\partial \Phi_\phi} &=& \sum_\beta [Y_\beta - \bar Y_\beta]\frac{\partial\bar Y_\beta}{\partial\Phi_\phi} = 0\nonumber\\
\frac{\partial \ln\cL}{\partial \Theta_\alpha} &=& \sum_\beta [Y_\beta - \bar Y_\beta]\frac{\partial\bar Y_\beta}{\partial\Theta_\alpha} =0.
\end{eqnarray}
Here we make an approximation that $\bar Y_\beta$ is linearly dependent on $\Theta$ and $\Phi$. This is accurate locally, but whether it will be a sufficiently accurate approximation globally depends on an interplay of parameter non-linearity and constraining power of the data. As shown in \citet{DALI} and \citet{DALII}, the approximation that $\bar Y_\beta$ is linearly dependent on $\Theta$ and $\Phi$ could be extended to higher-order terms, but if the data are sufficiently constraining, then those higher-order terms are rendered irrelevant as they only dominate in parameter regions that are assigned negligible posterior probability. Thus, our linear approximation is expected to also handle a wide range of non-linear parameters accurately, as long as the data are sufficiently constraining.

Expanding around the baseline ML point, where $\bar Y=\bar Y^0$ and $\Theta_\alpha=\Theta^0_\alpha; \Phi_\phi=0$, we introduce the shift of the old parameters 
\begin{equation}
\Delta\Theta_\alpha\equiv \Theta_\alpha-\Theta^0_\alpha. 
\end{equation}
This shift describes the displacement in the posterior of the baseline model due to having introduced new parameters. It gives the distance (and direction) between the old best-fit and the location of the new best-fit to be expected when introducing the parameters $\Phi$.
We then have
\begin{equation}
\bar Y_\beta = \bar Y_\beta^0 + \sum_{\phi=1}^s \bbf_\beta^T\cdot  \left(\Phi_\phi\frac{\partial \bmu}{\partial \Phi_\phi}\right) + \sum_{\alpha=1}^m \bbf_\beta^T\cdot  \left(\Delta\Theta_\alpha\frac{\partial \bmu}{\partial \Theta_\alpha}\right).
\end{equation}
Defining non-square matrices $\V$ and $\W$ with elements
\begin{equation}
V_{\phi\beta} \equiv \bbf_\beta^T\cdot \frac{\partial \bmu}{\partial \Phi_\phi}; \quad W_{\alpha\beta} \equiv \bbf_\beta^T\cdot \frac{\partial \bmu}{\partial \Theta_\alpha},
\end{equation}
we have (now with summation convention over ranges that are set by context)
\begin{eqnarray} 
0 &= & \frac{\partial \ln\cL}{\partial \Phi_\phi} =  \left[Y_\beta - \bar Y^0_\beta - \Phi_\rho \V_{\rho\beta}-\Delta\Theta_\alpha \W_{\alpha\beta}\right]\V_{\phi\beta},  \nonumber\\
0 &=& \frac{\partial \ln\cL}{\partial \Theta_\alpha} =  \left[Y_\beta - \bar Y^0_\beta - \Phi_\rho \V_{\rho\beta}-\Delta\Theta_\eta \W_{\eta\beta}\right]\W_{\alpha\beta} .
\end{eqnarray}
These equations have vector solutions
\begin{eqnarray}
\Delta\Theta &=& \X^{-1}[\W-(\W\V^T)(\V\V^T)^{-1}\V]\Delta Y,\nonumber\\
\Phi &=& \Z^{-1}[\V-(\V\W^T)(\W\W^T)^{-1}\W]\Delta Y,
\label{eq:PhiDtheta}
\end{eqnarray}
where $\Delta Y_\beta \equiv Y_\beta - \bar Y_\beta^0$ and 
\begin{eqnarray}
\X &\equiv& \W\W^T-\W\V^T(\V\V^T)^{-1}\V\W^T,\nonumber\\
\Z &\equiv& \V\V^T - \V \W^T(\W\W^T)^{-1}\W\V^T.
\end{eqnarray}
These calculations require only $O(N^2)$ operations per parameter.  Note that we can equally well do this linear model analysis with the uncompressed data, at the expense of computing $\C^{-1}$, and similar analytic expressions are reached.

The central result of this section is Eq.~(\ref{eq:PhiDtheta}). It describes by how much the best-fit of the old parameters $\Theta$ will shift, and in which direction. It also yields a direct analytical estimate of the values that the new parameters $\Phi$ will take.
Eq.~(\ref{eq:PhiDtheta}) therefore provides an analytical study of whether the extended model will improve upon the baseline theory.

The variances of the baseline model parameters, the specific extended model parameters, and their covariances, are obtained as a block matrix constructed from $\V$ and $\W$:
\begin{equation}
 -\frac{\partial^2\cL}{\partial\Theta_\alpha\partial\Theta_\beta}  = W_{\beta\eta}W_{\alpha\eta};
\label{eq:Hessian1}
\end{equation}
\begin{equation}
-\frac{\partial^2\cL}{\partial\Phi_\phi\partial\Phi_\psi}  = V_{\phi\eta}V_{\psi\eta};
\label{eq:Hessian2}
\end{equation}
\begin{equation}
-\frac{\partial^2\cL}{\partial\Theta_\alpha\partial\Phi_\phi}  = W_{\alpha\eta}V_{\phi\eta}.
\label{eq:Hessian3}
\end{equation}

One  advantage of this basis is that the data can be compressed, and the agnostic gPCA coefficients with the smallest $\Gamma$ values discarded.  We will see in the example below that this may be an effective way to retain an expanded set of compressed data (including the MOPED coefficients and a subset of the most discrepant gPCA modes), so that alternative models can be confronted very rapidly (albeit approximately) with the compressed data, without undertaking a full MCMC analysis.  In the test case considered later in Sect.~\ref{sect:GalaxyImages}, retaining 5\% of the gPCA modes will prove to be enough to show that that baseline model is inadequate.

\section {Example: galaxy images}
\label{sect:GalaxyImages}
\begin{figure}
\includegraphics[width=0.45 \textwidth]{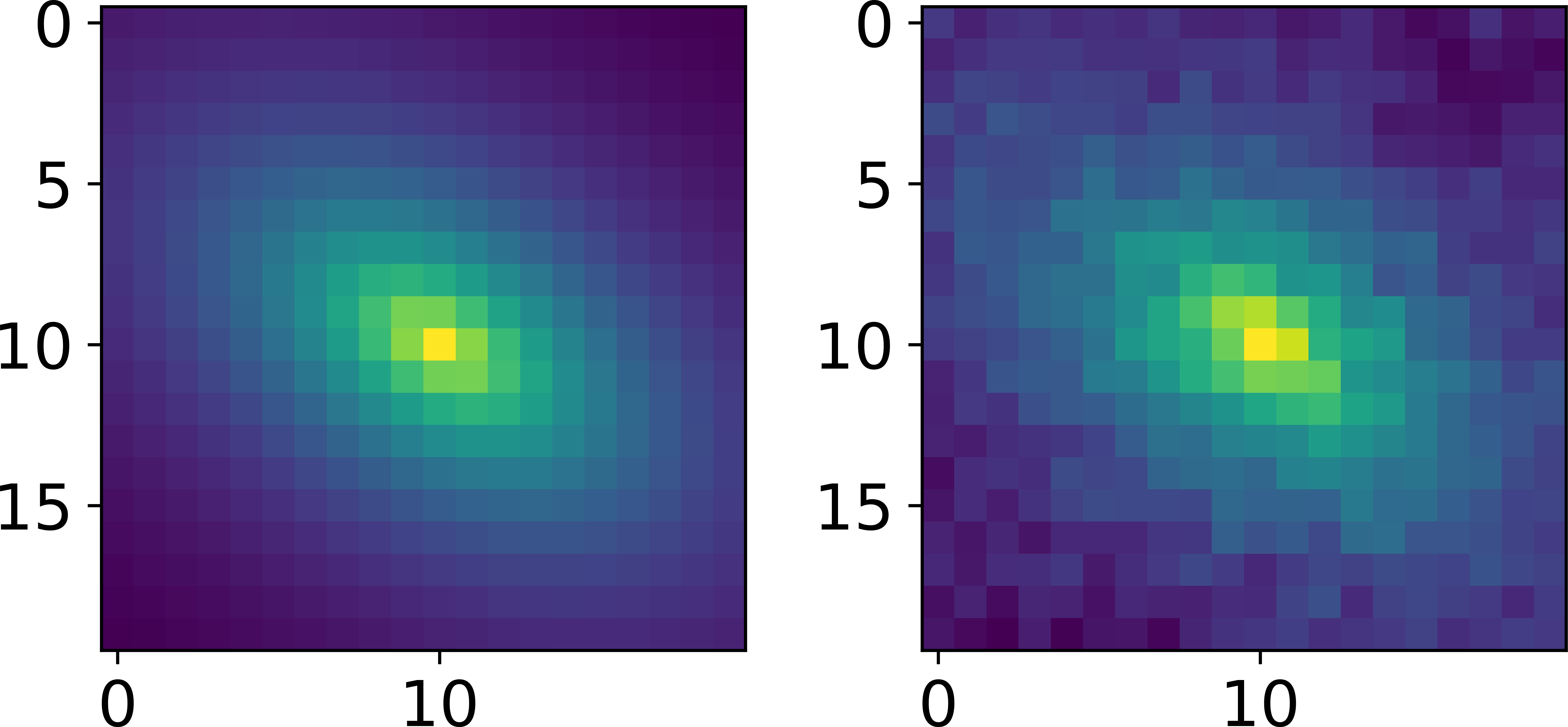}
\caption{Examples of the galaxy image model of Section~\ref{sect:GalaxyImages}. Left: Noise-free model of a pixelized galaxy. Right: the same galaxy with additive shot noise.} 
\label{fig:GalaxyImages20}
\end{figure}

The method is quite general, but for illustration, we consider the same 4-parameter model as in \cite{Heavens2017}, for the surface brightness of a thin exponential disk galaxy seen at an angle.  Parameters to be inferred are the scalelength $a$, central amplitude $A$, eccentricity $\epsilon$ and position angle $\phi$. The variables $r$ and $\varphi$ are polar coordinates of pixels in the image, meaning they are required to define the pixelized data, and are consequently not to be inferred.  We generate artificial data of such pixelized galaxy images and add noise which is assumed to be uncorrelated, with a white noise background uncertainty plus a source Poisson term.  The physical model of the brightness is then
\begin{equation}
\mu(r,\varphi; a,A,\epsilon,\phi)=A \exp\left[-\frac{r\sqrt{1+\epsilon^2-2\epsilon \cos 2(\varphi-\phi)}}{a\sqrt{1-\epsilon^2}}\right].
\label{model4}
\end{equation}
The observed surface brightness is 
\begin{equation}
    \hat{\mu}_{\rm obs}(r,\varphi) = \mu(r,\varphi; a,A,\epsilon,\phi) + n(r,\varphi).
\end{equation}
The Poisson shot noise $n(r,\varphi)$ is assumed to include enough photons to be accurately treated as Gaussian.

We generate an $n \times n$ galaxy image with parameters $a=n/4,\ A=100,\ \epsilon=0.25,\ \phi=1$.  We add a constant background level of 100. A visualization with $n=20$ is shown in Fig.~\ref{fig:GalaxyImages20}. 

\begin{figure}
    \centering
    \begin{minipage}{0.45\textwidth}
        \centering
        \includegraphics[width=0.9\textwidth]{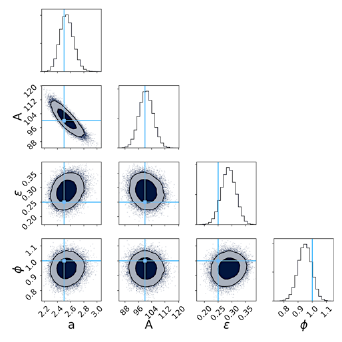} 
        \caption{Posterior constraints of the galaxy's four parameters from the full dataset of 100 data points.}
        \label{fig:PosteriorFull}
    \end{minipage}\hfill
    \begin{minipage}{0.45\textwidth}
        \centering
        \includegraphics[width=0.9\textwidth]{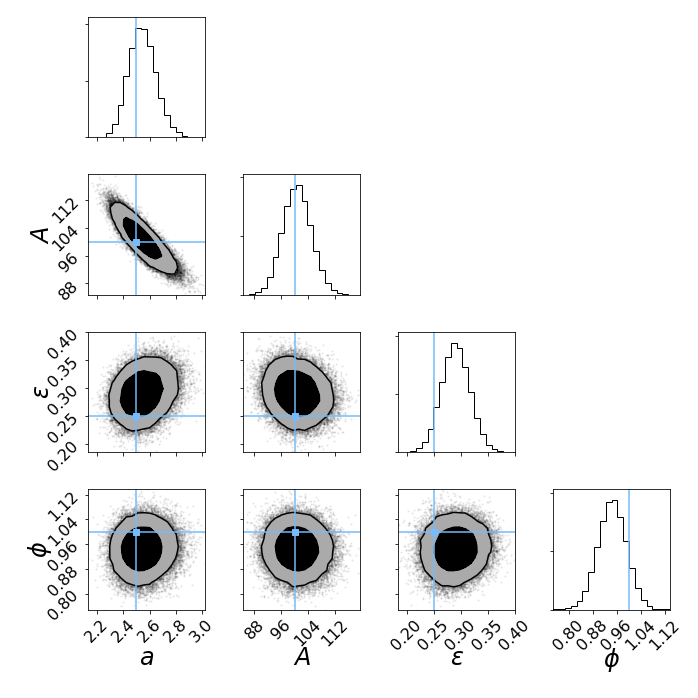} 
        \caption{Posterior from MOPED-compressed dataset, which comprises 4 data points only. The comparison with respect to Fig.~\ref{fig:PosteriorFull} reveals that the data compression indeed does not lose constraining power on the parameters. }
        \label{fig:PosteriorMOPED}
    \end{minipage}
\end{figure}

\begin{figure}
\includegraphics[width=0.47 \textwidth]{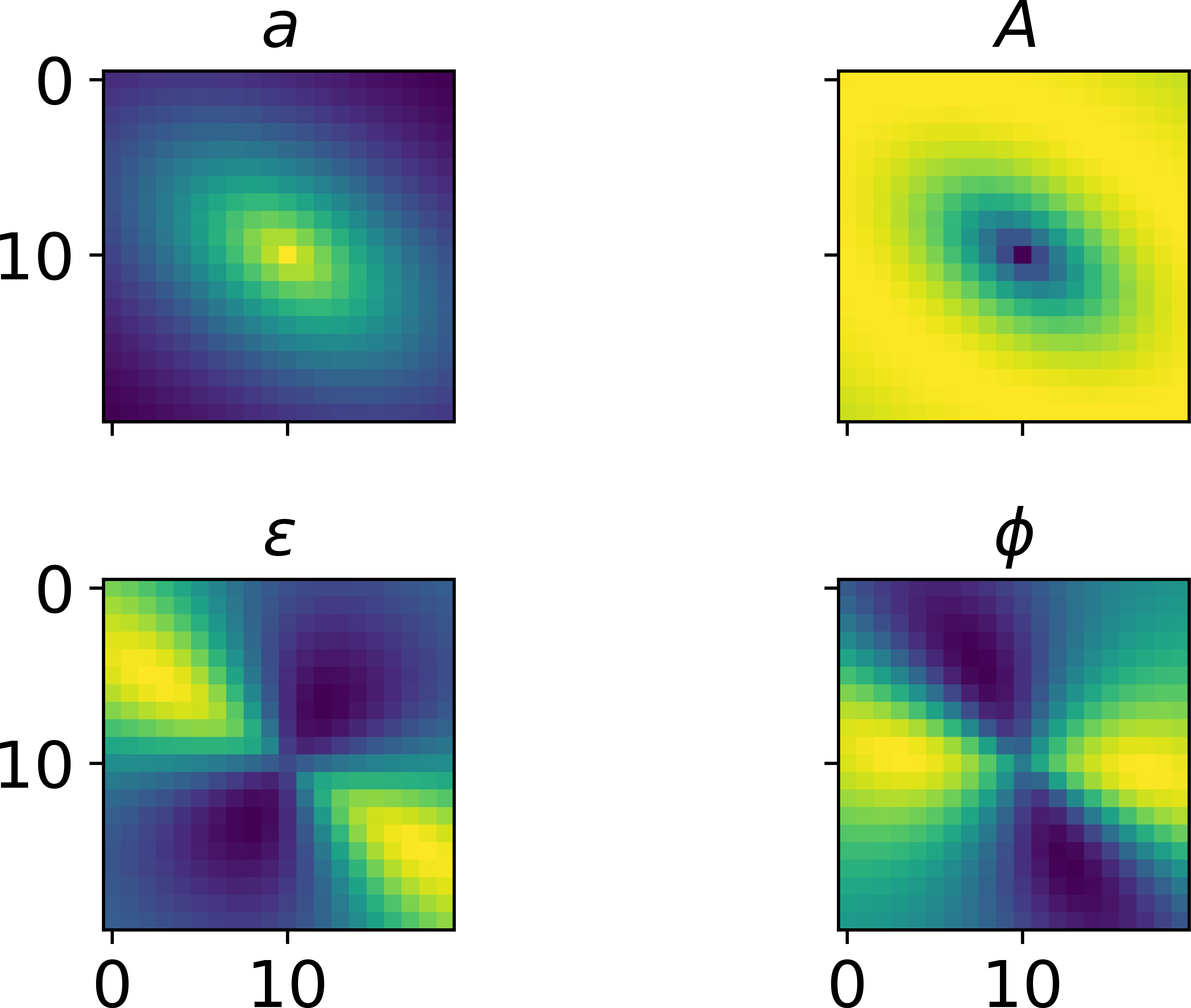}
\caption{The four MOPED filters, which filter the data for the information on the parameters $a,A,\epsilon,\phi$. The filters and the data are plotted in the pixel domain, using Cartesian coordinates. Applying the filters to the data implies taking the pixelwise dot product between these filters and the pixelized data. }
\label{fig:MOPEDvectors}
\end{figure}

The posterior using the full, uncompressed, dataset is shown in Fig.~\ref{fig:PosteriorFull}, for an image with $n=10$. The posterior constraints derived from the MOPED compressed data set are shown in Fig.~\ref{fig:PosteriorMOPED}, revealing that MOPED indeed does not lose precision, despite outputting only 4 compressed data points, and all parameters being non-linear except for $A$. The inference is done with Hamiltonian Monte Carlo sampling of 2 chains of 30000 points using Stan \citep{Stan}, with a burnin of 1000 points. The corner plots show posterior densities assuming a uniform prior, so likelihood and posterior will be used interchangeably in the parameter inference discussion. For model comparison purposes later we will assume a Gaussian prior.
 
\begin{figure}
\includegraphics[width=0.45\textwidth]{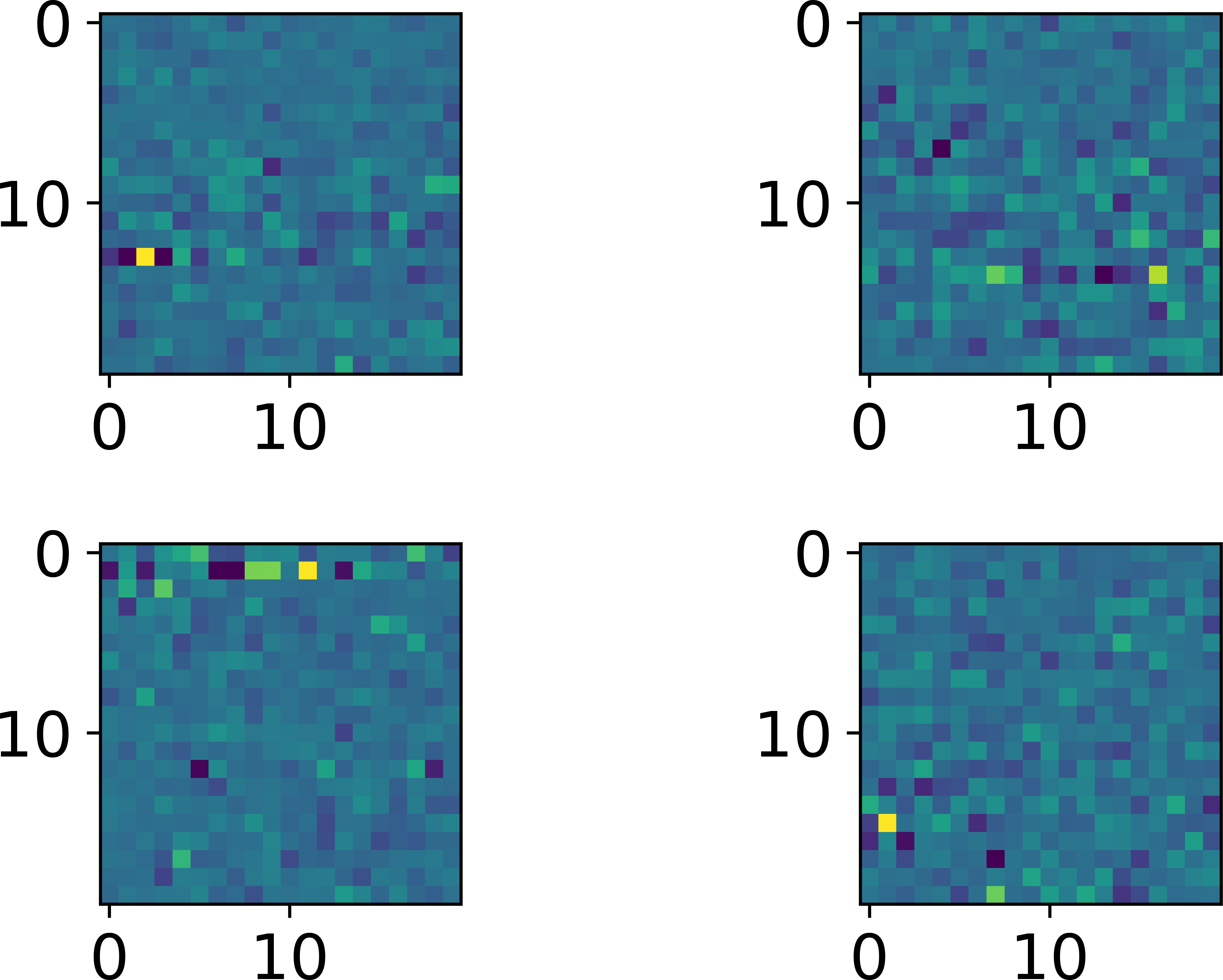}
\caption{Four gPCA vectors as used in the analysis. The gPCA vectors act as additional filters to sweep through the dataset, thereby outputting agnostic compressed data points from which non-standard physics can first be discovered and in a subsequent step constrained.}
\label{fig:PCAvectors}
\end{figure}

\begin{figure*}
\includegraphics[width=0.8\textwidth]{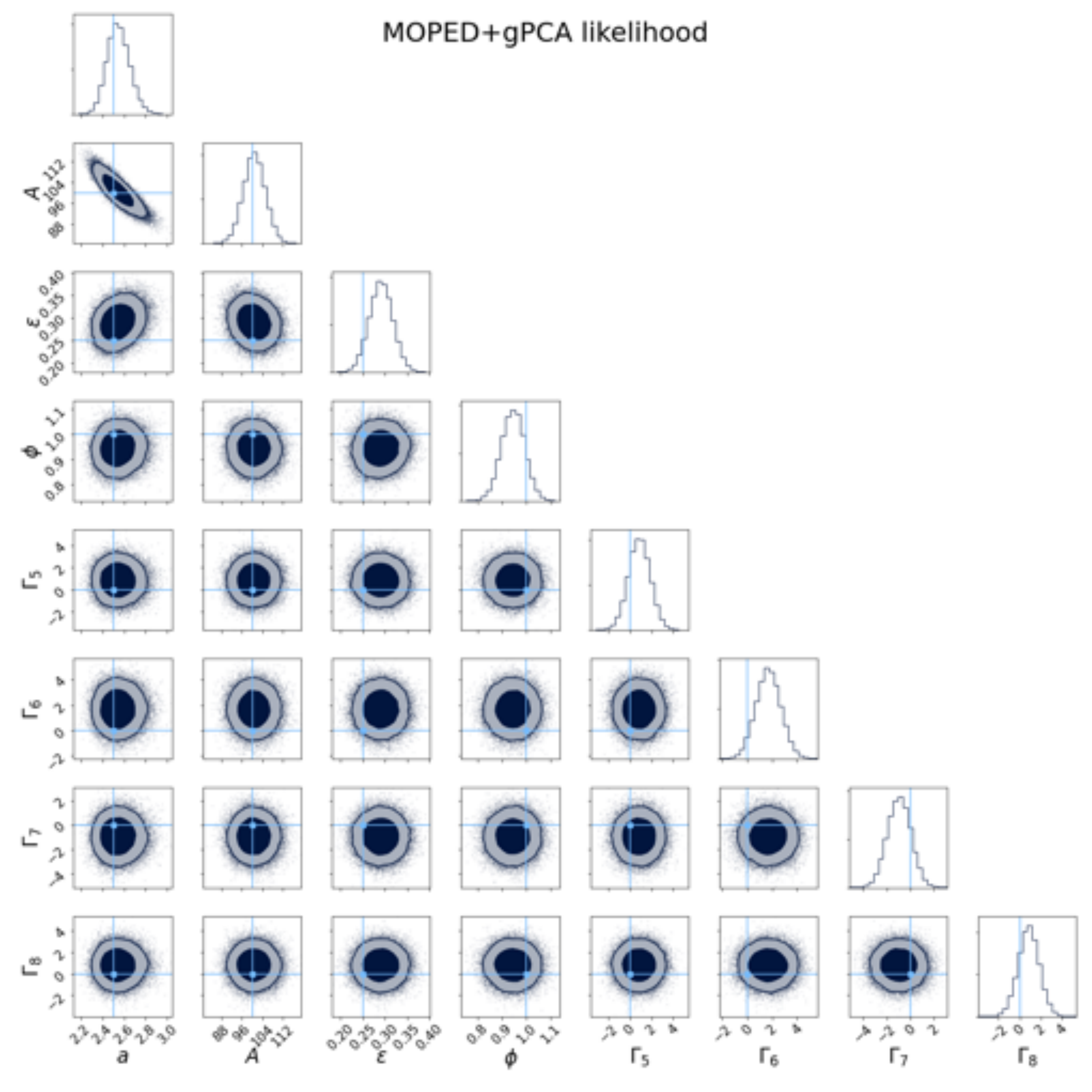}
\caption{Posterior from MOPED + 4 gPCA modes.  Note that the extra parameters are consistent with zero. The precision of the baseline model parameters is unchanged from analyzing the full data set, however many new parameters are analysed.}
\label{fig:MOPED_gPCA_4TrueTrueFalse}
\end{figure*}

The MOPED vectors are shown in Fig.~\ref{fig:MOPEDvectors}, and four gPCA vectors are shown in \ref{fig:PCAvectors}.  The inference with the MOPED coefficients plus the projections on to the first 4 gPCA modes is shown in Fig.~\ref{fig:MOPED_gPCA_4TrueTrueFalse}.  Note that the extra parameters are consistent with zero, implying the baseline model assumed by MOPED is sufficient to describe the data.  The posteriors are, within the sampling errors, unit width gaussians centred on the analytically-calculated means, as expected for a correctly-specified model expanded around the posterior mode.
With all additional components, the Bayesian evidence favours the baseline model with $\ln B_{\rm BE} = 53.7$, which is highly decisive in favour of the simple baseline model.  For information only, $\chi^2 = 107$ for 96 extra components.  

\subsection{Conservation of precision despite extended parameter set}

The means and standard deviations from MCMC analysis of the full likelihood and the compressed data are shown in Table~\ref{tab:results}.  The MOPED coefficients are derived assuming the correct model parameters. The maximum 96 extra gPCA components are added.  MOPED on its own incurs no information loss at Fisher level, and we see that this is still the case while jointly inferring the extra gPCA components.  This is expected, since the gPCA components are designed to have probability distributions that are independent of the MOPED coefficients.
\begin{table}
\begin{tabular}{l|lllll}
 & True & Mean  & Mean  & s.d. & s.d.  \\
& value & (full data) & (MOPED &  (full) & (MOPED\\
& & & +gPCA) & & +gPCA) \\\hline
 a & 2.5 & 2.54 & 2.54  & 0.10 & 0.10 \\
 A & 100 & 100.7&  100.7 &  4.3 & 4.3 \\
 $\epsilon$ & 0.25 & 0.29 & 0.29 &  0.03 & 0.03 \\
 $\phi$ & 1.0 & 0.94 & 0.95 & 0.05 & 0.05 
\end{tabular}
\caption{Mean and standard deviation from chains run with the full dataset (100 data points), and with the MOPED+gPCA data. By construction, even with the MCMC runs being different, MOPED+gPCA maintains the optimal constraints of MOPED despite having introduced 96 additional parameters to fit for. The peak positions are maintained as well.}
\label{tab:results}
\end{table}

\subsection{Wrong physical model: presence of a bulge}
\label{sect:wrongbulge}

 To test the ability of the method to detect failures of the baseline model, we generate data from another model. In this case, we analyse the data with the full set of baseline modes, and we might expect to see large values of the gPCA coefficients $\Gamma_\gamma$, and a Bayes factor that prefers their inclusion in the agnostic model. 

\begin{figure}
\includegraphics[width=0.45 \textwidth]{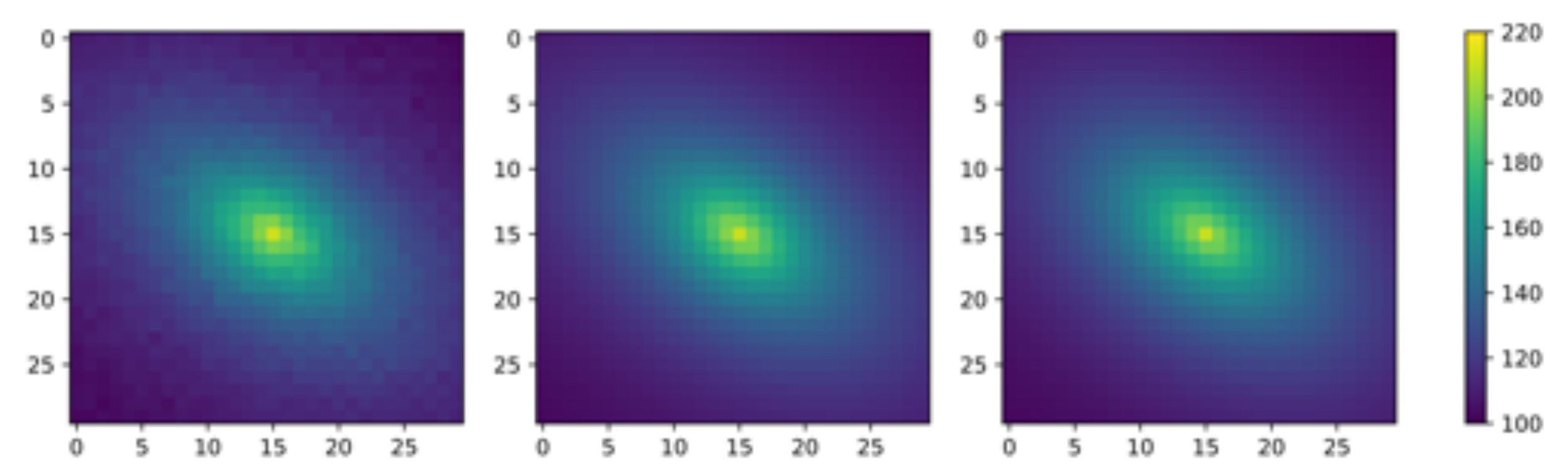}
\caption{Image generated from the extended Disk+bulge full data model with (left) and without noise (centre). On the right is the maximum likelihood baseline (disk without bulge) model, which visually matches the data quite closely but yields incorrect baseline parameters.
}
\label{fig:BulgeImage}
\end{figure}

Specifically, we generate an image which has an additional circularly symmetric bulge with a Gaussian projected radial profile, parameterized as an additive term $A_{\rm b} \exp(-\alpha_{\rm b} r^2)$.  An example
image is shown in Fig.~\ref{fig:BulgeImage} where we show a noisy disk+bulge image, the noise-free model, and the maximum likelihood solution that assumes the baseline model (disk, with no bulge). The latter solution is an adequate fit by eye, but the baseline-model parameters are not correctly recovered.

\begin{figure*}
\includegraphics[width=0.8 \textwidth]{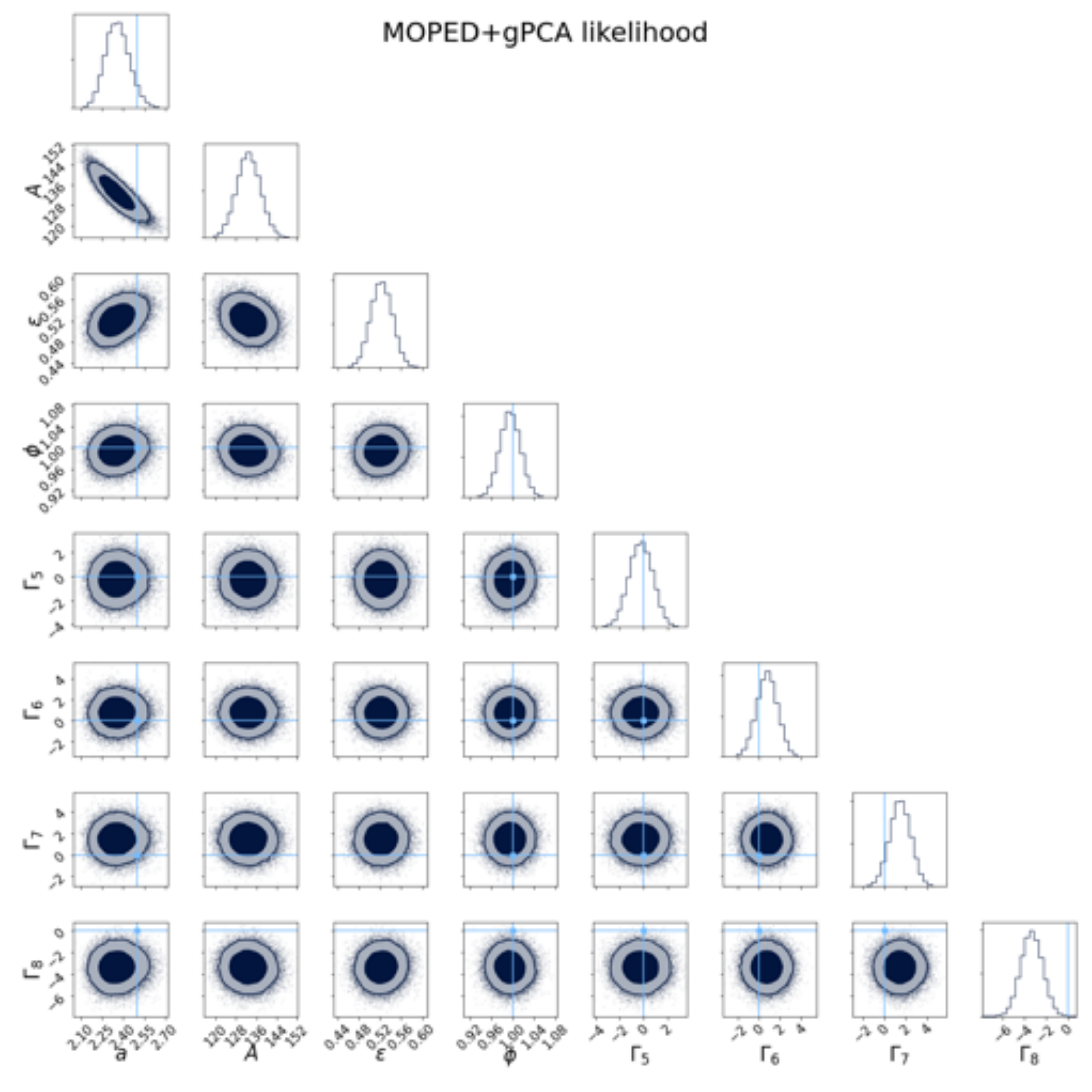}
\caption{MOPED + 4 gPCA inferred parameters for data that are drawn from a disk+bulge model, whereas the baseline physical model has disk only (see text for details). Note for example that the distribution of $\Gamma_8$ is noticeably offset from zero. }
\label{fig:MOPEDPCAchain4Bulge}
\end{figure*}

Fig.~\ref{fig:MOPEDPCAchain4Bulge} shows the posterior for the physical parameters and 4 gPCA coefficients assuming a uniform prior, for a galaxy with $A_{\rm b}=0.3A$, $\alpha_{\rm b}=1/(2a^2)$ and $n=10$.  Note that the posterior probability density that $\Gamma_8=0$ in particular is very low.  Bayesian Evidence disfavours the physical model, with $\ln B_{\rm BE} = -6.8$, for a prior width $\Sigma_\gamma=5$ for all additional gPCA components, showing strong evidence for the need to modify the data model.  As is always the case, the Bayes factor depends on the prior: with $\Sigma_\gamma=10$, $\ln B_{\rm BE} = -9.7$, with the same conclusion. For information only, $\chi^2=13.8$ for the 4 gPCA modes. With more components, the evidence would be stronger.

\subsection{Finding a better suited model}
\label{sect:bettermodel}

The agnostic extended (gPCA) modes can be used to confront any  physical model. Here we consider a nested model, where the baseline model is a special case of the extended model. By projecting the derivatives in the extended model onto the gPCA modes, we can infer the extended parameters, and correct the baseline parameters, using equation (\ref{eq:PhiDtheta}).  We consider the additive Gaussian bulge described in Sec.~\ref{sect:wrongbulge}. We infer the central amplitude $\Phi_0 = A_{\rm b}$ and the scale parameter $\Phi_1=\alpha_{\rm b}$, which is highly nonlinear and more challenging to infer.  For this reason, we use larger images with $n=30$.  In this example, $A_{\rm b}=0.2A$ and  $\alpha_{\rm b}=1/18$, corresponding to a Gaussian width of 3 pixels.

\begin{figure}
\includegraphics[width=0.45 \textwidth]{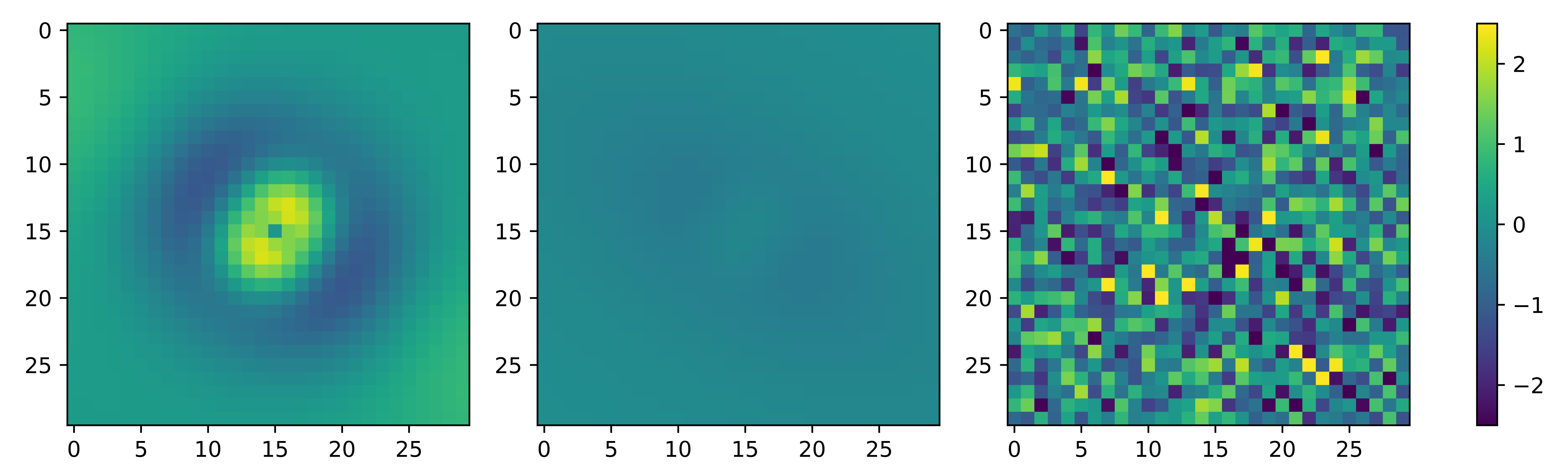}
\caption{Difference between maximum likelihood baseline model fit and the image generated from the extended model. Centre panel shows the differences after the linear correction of the baseline model parameters and the bulge amplitude (without noise).  The right panel shows the difference between the data and the extended model inferred using the MOPED + gPCA modes.}
\label{fig:DiffImage}
\end{figure}

\begin{figure}
\includegraphics[width=0.45 \textwidth]{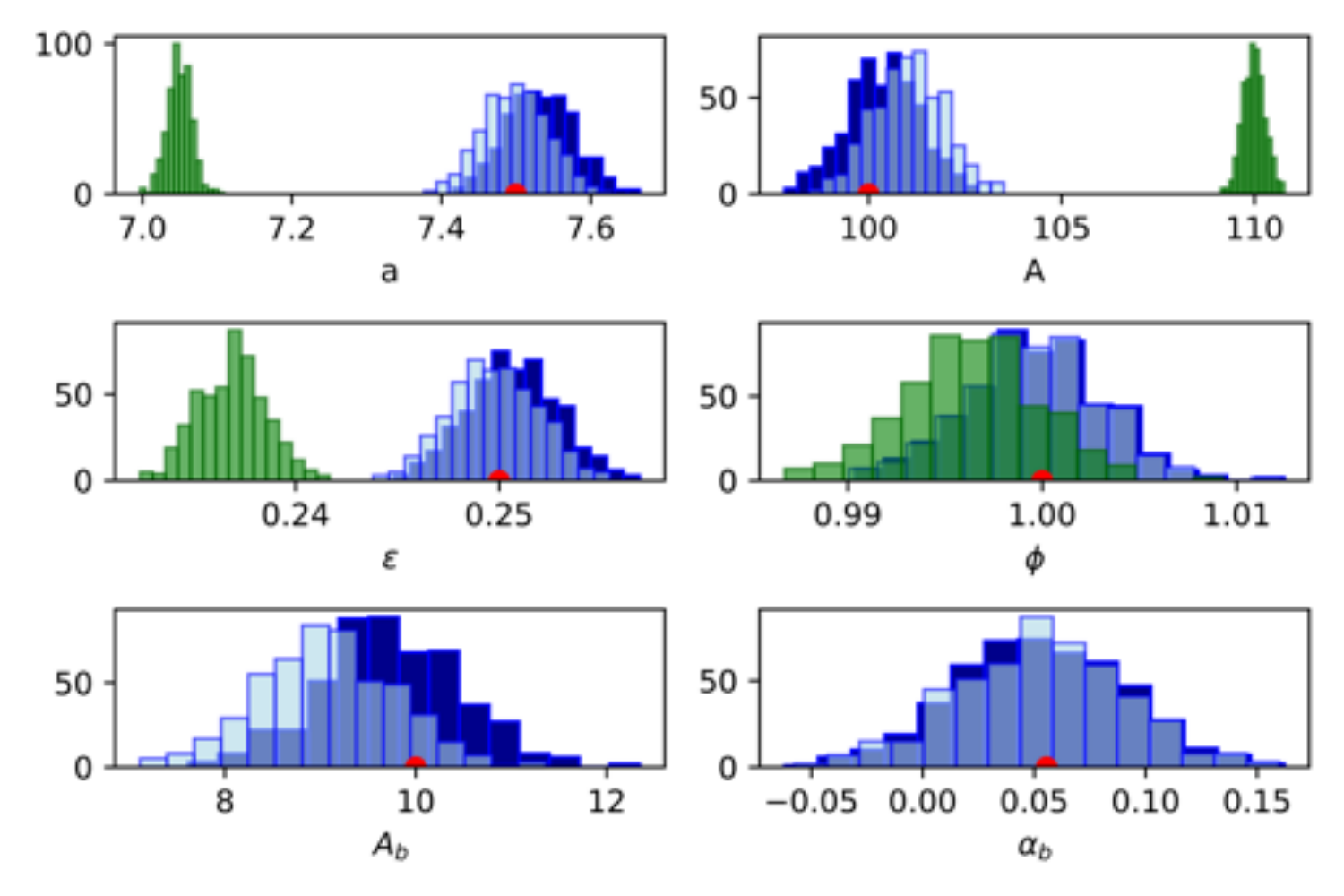}
\caption{Maximum likelihood estimates of the bulge amplitude $A_{\rm b}$from 500 noise realisations, and of the baseline model parameters, using all the gPCA modes (dark blue), and using half of the modes with the highest values of $|\Delta Y|$, which are most discrepant with the baseline model expectation (light blue).  True values are marked with the red dot, and the distributions of the baseline parameters from the baseline MOPED analysis are shown in green.}
\label{fig:MLestimates50}
\end{figure}

 Fig.~\ref{fig:DiffImage} shows the difference between the true extended model and the maximum-likelihood baseline fit, showing that the baseline model fails to reproduce the central structure with discrepancies of the order of 1.2\%.  In the middle panel, we see the results of using the gPCA modes to constrain the extended model, calculating shifts to the baseline parameters and estimating the extended model parameter (the bulge amplitude).  These are computed according to equation (\ref{eq:PhiDtheta}).  The residuals are now at the 0.2\% level.  Fig.~\ref{fig:MLestimates50} shows in dark blue the distribution of recovered maximum likelihood baseline and extended model parameters, with the truth shown as a red circle.  For this application, the linear assumption recovers the parameters reasonably well.  The pale blue histogram shows that the recovery is still reasonably good when only the most discrepant 50\% of the gPCA modes is used. The standard deviations of all the dark blue histograms agree to better than 4\% with the posterior widths predicted in eqs. (\ref{eq:Hessian1}-\ref{eq:Hessian3}), consistent with the variation expected from the finite number of samples, but note that they are not quite the same quantity.

\begin{figure}
\includegraphics[width=0.45 \textwidth]{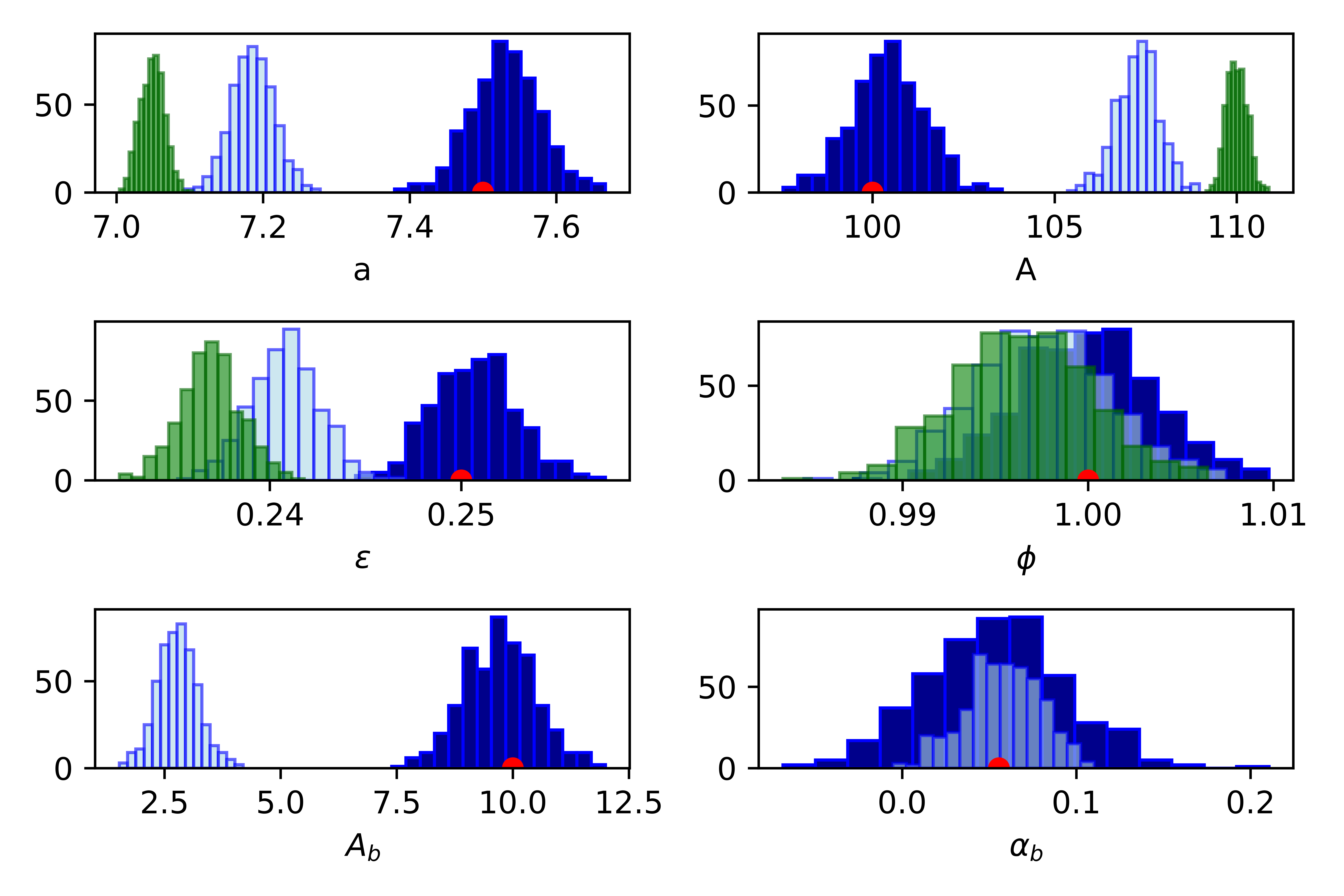}
\caption{As in Fig.~\ref{fig:MLestimates50}, but with the pale blue histograms showing the ML estimates from retaining only 5\% of the gPCA modes.  This radical compression does not recover the parameters well, but still provides evidence (especially from $A_{\rm b}\ne 0$)that the baseline model is not preferred over the extended model, which should be investigated further.}
\label{fig:MLestimates95}
\end{figure}

\begin{figure}
\includegraphics[width=0.45 \textwidth]{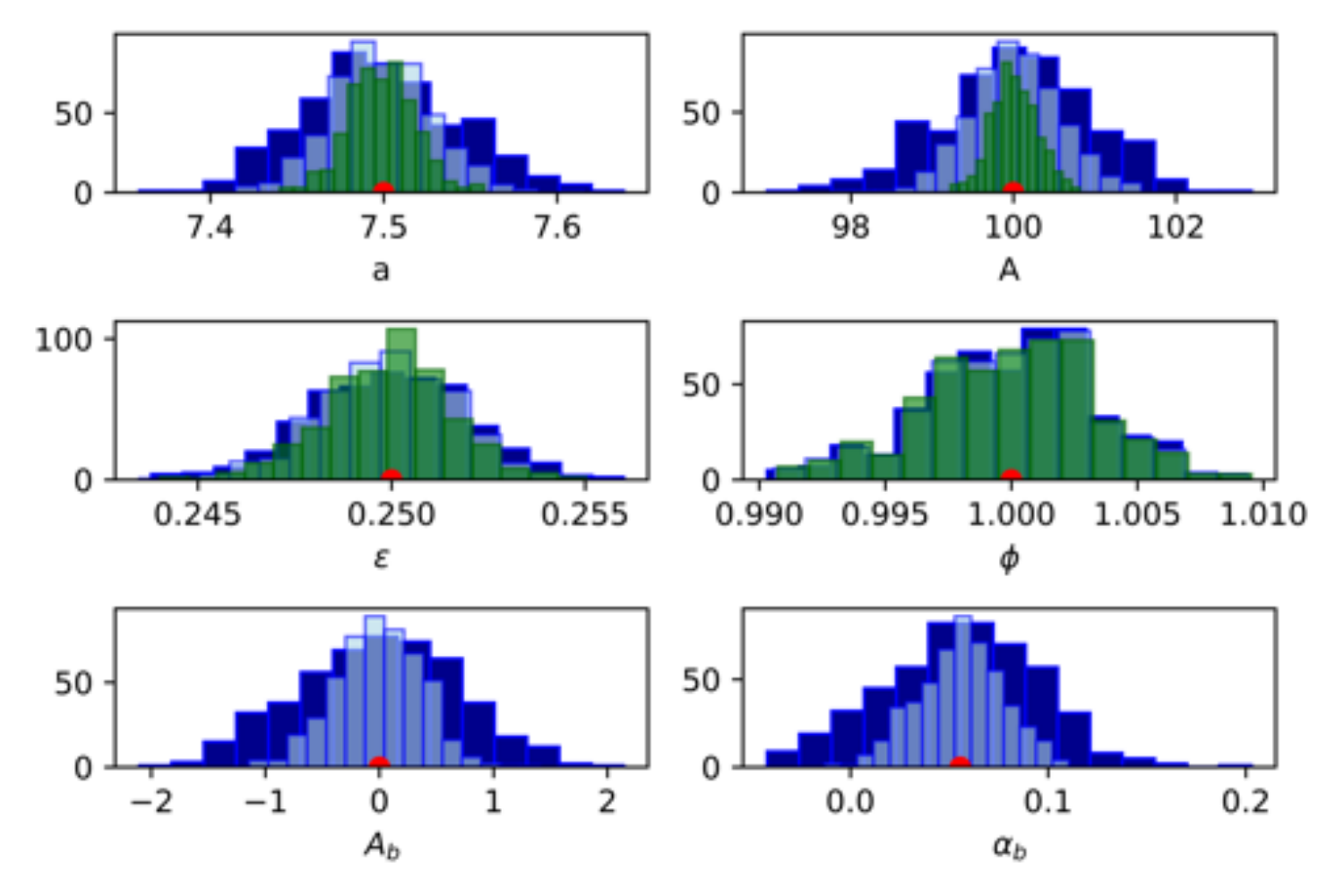}
\caption{As in Fig.~\ref{fig:MLestimates95}, but here there is no bulge ($A_{\rm b}=0$), and the baseline model is correct. The highly compressed dataset of 4 MOPED coefficients plus 5\% of the gPCA modes is consistent with the baseline model, as desired.}
\label{fig:MLestimatesNoBulge}
\end{figure}

If we reduce the proportion of gPCA modes further, the recovery of parameters becomes poor, but a radical compression can still be useful.  Fig.~\ref{fig:MLestimates95} shows the recovered parameters from only 5\% of the gPCA modes, plus the MOPED coefficients.  The recovery is poor, but still shows clear evidence (especially in $A_{\rm b}$) that the baseline model is inadequate, and extended models should be considered more carefully.  The moral of this part is that a massive data compression to the MOPED coefficients plus a few gPCA modes may be used to test at a basic level the baseline model against various extended models, without further MCMC calculation, and using only fast linear algebra.  Finally, for completeness, we show in Fig.~\ref{fig:MLestimatesNoBulge} that if the baseline model is correct, and there is no bulge, then these 5\% most discrepant modes will support the baseline model, and do not erroneously show  evidence that an extended model is preferred.

\section{Conclusions}
\label{sect:conclusions}
In this paper we have extended the MOPED extreme data compression algorithm to address a specific issue, that the MOPED algorithm is inextricably tied to a particular data model, which we describe as the baseline theoretical model.  In the original MOPED compression setup, this has the disadvantage that if a different theory is considered, the analysis has to be redone from scratch.  

In this paper, we augment the MOPED weighting vectors with a set of orthogonal modes that span the subspace that is not in the local hyperplane that is spanned by the MOPED vectors.  The construction of these modes leads to a generalised principle component analysis (gPCA) of the subspace, and the combined set of MOPED-PC filters have a number of desirable features.  Between them, they correspond to all possible directions in data space, and each mode has connected with it an amplitude whose expected distribution is a unit variance Gaussian centred on zero, if the baseline model is correct.  These modes are independent of each other and of the MOPED coefficients, if the fiducial parameters used for the compression are correct. 

By including all the additional modes, we can investigate a `generic extended model' with  The advantages of these properties are several:  firstly, the maximum likelihood amplitude of the modes can be calculated analytically once the baseline model has been investigated with MCMC - further (higher-dimension) MCMC analyses are not necessary; secondly, because of the independence, the investigation of these extra modes has no bearing on the precision of the baseline parameters, whose inference is not degraded; thirdly, by assigning priors to the extended mode amplitudes, the Bayes factor between the baseline model and the generic extended model can be computed analytically, so the baseline model can be compared with a generic model to see if the data indicate that further models should be investigated.  

The extended gPCA modes also form a convenient basis for confronting the data with specific extended models, whose data derivatives can be projected onto the modes, and the extra parameters of the extended theory determined under the approximation that the changes in the mean data vector are linear in the parameters.  If the extended model holds, then the posterior distributions of the baseline parameters from the MOPED analysis will be systematically wrong, but they can be corrected, and the extended model parameters determined, using linear algebra alone.  This is not a surprise for linear models, and the effectiveness of the procedure will depend on how good the linear approximation is.

Finally, there is the option to keep the radical data compression of MOPED, which is optimal for the baseline model, and increase the size of the compressed dataset by retaining the gPCA modes that are most discrepant with the baseline theory.  This is a sensible way to maximise sensitivity to new physics that might come along later.  We find in the example case considered that a reduction of the gPCA modes to only 5\% of the original size can give a very strong indication that the baseline model needs extension, even if these modes alone are insufficient to correct fully the baseline parameters and to recover the true extended parameters.

\subsection*{Acknowledgements}

We thank the anonymous referee for helpful comments.

\subsection*{Data Availability}

No new data were generated or analysed in support of this research.

\label{lastpage}
\end{document}